\documentclass[published]{JHEP3}

\JHEP{00(2005)000}

\JHEPspecialurl{http://jhep.sissa.it/JOURNAL/JHEP3.tar.gz}

\usepackage{amssymb}
\usepackage{latexsym}
\usepackage{mathptm}
\usepackage{epsfig}

\def\beq{\begin{equation}}
\def\eeq{\end{equation}}
\def\beqn{\begin{eqnarray}}
\def\eeqn{\end{eqnarray}}

\title{Black Hole Remnants at the LHC}
\author{
\normalsize{Benjamin Koch${}^{1,3}$\thanks{koch@th.physik.uni-frankfurt.de},
Marcus Bleicher${}^1$\thanks{bleicher@th.physik.uni-frankfurt.de} and
Sabine Hossenfelder${}^2$\thanks{sabine@physics.arizona.edu}
}\\
\vspace*{-.1cm}\small{${}^1$ Institut f\"ur Theoretische Physik}\\
\vspace*{-.1cm}\small{J. W. Goethe-Universit\"at}\\
\vspace*{-.1cm}\small{Max von Laue Strasse 1}\\
\vspace*{.1cm}\small{60438 Frankfurt am Main, Germany}\\
\vspace*{-.1cm}\small{${}^2$Department of Physics}\\
\vspace*{-.1cm}\small{University of Arizona}\\
\vspace*{-.1cm}\small{1118 East 4th Street}\\
\vspace*{.1cm}\small{Tucson, AZ 85721, USA}\\
\vspace*{-.1cm}\small{${}^3$Frankfurt International Graduate School for Science (FIGSS)}\\
\vspace*{-.1cm}\small{J. W. Goethe-Universit\"at}\\
\vspace*{-.1cm}\small{Max von Laue Strasse 1}\\
\vspace*{-.1cm}\small{60438 Frankfurt am Main, Germany} }
\bigskip
\date{}
\received{July 15, 2005}            
\revised{}
\accepted{}            

\abstract{
Within the scenario of large extra dimensions, the Planck scale is
lowered to values soon accessible.  Among the predicted effects, the
production of TeV mass black holes at the {\sc LHC} is one of the
most exciting possibilities. Though the final phases of the black
hole's evaporation are still unknown, the formation of a black hole
remnant is a theoretically well motivated expectation. We analyze
the observables emerging from a black hole evaporation with a
remnant instead of a final decay. We show that the formation of a
black hole remnant yields a signature  which differs substantially
from a final decay. We find the total transverse momentum of the
black hole event to be significantly dominated by the presence of a
remnant mass providing a strong experimental signature for black
hole remnant formation.
}

\begin{document}
\section{Introduction}

High energetic particle collisions will eventually lead to strong
gravitational interactions and result in the formation of a black
hole's horizon. In the presence of large additional compactified
dimensions \cite{Arkani-Hamed:1998rs}, it could be possible that the
threshold for black hole production lies within the accessible range
for future experiments (e.g. LHC, CLIC). In the context of
models with such large extra dimensions, black hole production is
predicted to drastically change high energy physics already at the
LHC. These effective models with extra dimensions are
string-inspired
\cite{Antoniadis:1990ew,Antoniadis:1996hk,Dienes:1998vg} extensions
to the Standard Model in the overlap region between 'top-down' and
'bottom-up' approaches.

The possible production of
TeV-scale black holes at the {\sc LHC} is
surely one of the most exciting predictions of physics beyond
the Standard Model and has
received a great amount of interest during the last years
\cite{adm,Banks:1999gd,ehm,dim,Giddings3,Mocioiu:2003gi,Voloshin:2001fe,Giddings:2001ih,Solodukhin:2002ui,Jevicki:2002fq,GrWav,Polchi,Formfactors,bhsonst,Rychkov:2004sf,Cardoso:2005jq}.
For reviews on the subject the interested reader is referred to \cite{Kanti:2004nr}.

Due to their Hawking-radiation \cite{Hawk1}, these small black holes
will have a temperature of some $100$~GeV and will decay quickly
into $\sim 10-25$ thermally distributed particles of the Standard
Model (before fragmentation of the emitted partons). This yields a
signature unlike all other new predicted effects. The black hole's
evaporation process connects quantum gravity with quantum field
theory and particle physics, and is a promising way towards the
understanding of Planck scale physics.

Thus, black holes are a fascinating field of research which features
an interplay between General Relativity, thermodynamics, quantum
field theory, and  particle physics. The investigation
of black holes  would allow to test Planck scale effects and the
onset of quantum gravity. Therefore, the understanding of the black
holes properties is a key knowledge to the phenomenology of physics
beyond the Standard Model.

Recently, the production of black holes has been incorporated into
detailed numerical simulations for  black hole production and decay
in ultra-high energetic hadron-hadron interactions
\cite{CHARYBDIS,Atlas,BHex}.

So far the numerical simulation has assumed that the black hole
decays in its final phase completely into some few particles of the
Standard Model. However, from the theoretical point of view, there
are strong indications that the black hole does not evaporate
completely, but leaves a stable black hole remnant. In this work, we
will include this possibility into the numerical simulation and
examine the consequences for the observables of the black hole
event. These investigations might allow  to reconstruct initial
parameters of the model from observed data and can shed light onto
these important questions.

The aim of this investigation is not to derive the formation of a
remnant from a theory of modified gravity but to incorporate the
assumption of such a remnant into the possible signatures  black
hole events in high energetic particle interactions.

This paper is organized as follows: the next section briefly reviews
basic facts about black holes in extra dimensions. Section \ref{BHr}
discusses the issue of black hole remnants and introduces a useful
parametrization for the thermodynamical treatment. In section
\ref{signatures}, we discuss the results of the numerical
simulation. We conclude in section \ref{Concl}.

Throughout this paper we adopt the convention $\hbar=c=k_B=1$.

\section{Black Holes in Extra Dimensions}
\label{BHinXD}

Arkani-Hamed, Dimopoulos and Dvali \cite{Arkani-Hamed:1998rs}
proposed a solution to the hierarchy problem by the introduction of
$d$ additional compactified space-like dimensions in which only
gravitons can propagate. The Standard Model (SM) particles are bound
to our 4-dimensional sub-manifold, called our 3-brane.

Gauss' law then relates the fundamental mass scale of the extended
theory, $M_{\rm f}$, to the apparent Planck scale, $m_{\rm p}\sim
10^{16}$~TeV, by the volume of the extra dimensions. In the case of
toroidal compactification on radii of equal size this yields \beqn
m_{\rm p}^2 = M_{\rm f}^{d+2} R^d \quad. \eeqn Thus, for large
radii, the Planck scale can be lowered to a new fundamental scale,
$M_{\rm f}$ which can lie close by the electroweak scale.

The radius $R$ of the extra dimensions is then in the range mm to
$10^3$~fm for $d$ from $2$ to $7$, or the inverse radius $1/R$ lies
in energy range eV to MeV, resp. Since this radii are large compared
to the Planck scale, this setting  is called a scenario with large
extra dimensions ({\sc LXD}s). For recent constraints on the
parameter of the model see e.g. \cite{Cheung:2004ab}.

Using the higher dimensional Schwarzschild-metric \cite{my}, it can be derived that the
horizon radius $R_H$ of a black hole is substantially increased in the presence of
{\sc LXD}s \cite{Banks:1999gd}, reflecting the fact
that gravity at small distances becomes stronger. For a black hole of mass $M$ one finds
\begin{eqnarray} \label{ssr}
R^{d+1}_H= \frac{1}{d+1} \frac{1}{M_{\rm f}^{d+1}} \frac{M}{M_{\rm f}}\quad.
\end{eqnarray}
The horizon radius for a black hole with mass $\approx$~TeV is then $\approx 10^{-3}$~fm, and thus
$R_H \ll R$ for black holes which can possibly be produced at colliders or in ultra high energetic
cosmic rays. Also in higher dimensions the entropy, $S$, of the black hole is
proportional to its horizon surface which is given by
\begin{eqnarray}
{\cal{A}}_{(d+3)} &=& \Omega_{(d+3)} R_H^{d+2}
  \label{oberfl}
\end{eqnarray}
where $\Omega_{(d+3)}$ is the surface of the $d+3$-dimensional unit sphere
\begin{eqnarray}
\Omega_{(d+3)} = \frac{2 \pi^{\frac{d+3}{2}}}{\Gamma({\frac{d+3}{2}})}\quad.
\end{eqnarray}

Black holes with masses in the range of the lowered Planck scale
should be a subject of quantum gravity. Since there is yet no theory
available to perform these calculations, the black holes are treated
as semi classical objects which form intermediate meta-stable states.
Thus, the black holes are produced and decay according to the semi
classical formalism of black hole physics.

To compute the production probability, the cross-section of the
black holes can be approximated by the classical geometric
cross-section \cite{dim,Giddings3}
\begin{eqnarray} \label{cross}
\sigma(M)\approx \pi R_H^2 ~ \Theta (M-M_{{\rm min}})\quad,
\end{eqnarray}
an expression which contains only  the fundamental Planck scale as coupling constant. $M_{\rm min}$ is
the threshold above which the production can occur and expected to be a few $\times M_{\rm f}$.
As has been shown recently \cite{Rizzo:2005fz}, such a threshold arises naturally in certain types
of higher order curvature gravity.

The semi classical black hole cross section \cite{Banks:1999gd,dim,Giddings3}
has been under debate
\cite{Voloshin:2001fe}, but further investigations justify the use
of the classical limit at least up to energies of $\sim 10 M_{\rm
f}$ \cite{Giddings:2001ih,Solodukhin:2002ui,Jevicki:2002fq}. It has further been
shown that the naively expected classical result remains valid also
in string-theory \cite{Polchi}. However, this interesting topic is
still a matter of ongoing research, see e.g. the very recent
contributions in Refs. \cite{Rychkov:2004sf}.

A common approach to improve the naive picture of colliding point particles,
is to treat the creation of the horizon as a
collision of two shock fronts in an Aichelburg-Sexl geometry describing
the fast moving particles
 \cite{GrWav}. Due to the high velocity of the moving particles, space time
before and after the shocks is almost flat and the geometry
can be examined for the occurrence of trapped surfaces.

These semi classical considerations do also give rise to form
factors which take into account that not the whole initial energy is
captured behind the horizon. These factors have been calculated in
\cite{Formfactors} and depend on the number of extra dimensions,
however their numerical values are of order one. Setting $M_{\rm
f}\sim 1$TeV and $d=2$ one finds $\sigma \sim 1$~TeV$^{-2}\sim
400$~pb. With this cross section it is further found that these
black holes will be produced at the {\sc LHC} in huge numbers on the order
of  $\sim 10^9$ per year \cite{dim}.

Once produced, the black holes will undergo an evaporation process whose thermal
properties carry information about the parameters $M_{\rm f}$ and $d$. An analysis of the evaporation
will therefore offer the possibility to extract knowledge about the topology of our
space time and the underlying theory.

The evaporation process can be categorized in three characteristic stages \cite{Giddings3}:
\begin{enumerate}
\item {\sc Balding phase:} In this phase the black hole radiates away the multi-pole moments it has inherited
from the initial configuration, and settles down in
a hairless state. During this stage, a certain fraction of the initial mass will be lost in gravitational
radiation.

\item {\sc Evaporation phase:} The evaporation phase starts with a spin down phase in which the Hawking
radiation \cite{Hawk1} carries away the angular momentum, after which it proceeds with the emission of thermally
distributed quanta until the black hole reaches Planck mass. The radiation spectrum contains
all Standard Model particles, which are emitted on our brane, as well as gravitons, which are also
emitted into the extra dimensions. It is expected that most of the initial
energy is emitted during this phase in Standard Model particles \cite{ehm}.

\item {\sc Planck phase:} Once the black hole has reached a mass close to the Planck mass, it falls into
the regime of quantum gravity and predictions become increasingly difficult. It is generally
assumed that the black hole will either completely decay in some last few Standard Model particles or
a stable remnant will be left, which carries away the remaining energy.
\end{enumerate}

The evaporation phase is expected to be the most important phase for
high energy collisions. The characteristics of the black hole's evaporation in this phase can be
computed using the laws of black hole thermodynamics and are obtained by
first solving the field equations for the metric of the black hole,
then deriving the surface gravity, $\kappa$, from which the temperature of the black hole follows via
\begin{eqnarray} \label{kappa}
T = \frac{\kappa}{2 \pi} \quad.
\end{eqnarray}
By identifying the total energy of the
system with the mass of the black hole one then finds the entropy $S$ by integrating the
thermodynamical identity
\begin{eqnarray}
\frac{\partial S}{\partial M} = \frac{1}{T} \quad,
\end{eqnarray}
which fixes the constant factor relating the entropy to the horizon surface. A possible additive
constant is generally chosen such that the entropy is zero for vanishing horizon surface. In contrast
to classical thermodynamical objects, this does not necessarily imply that the entropy vanishes
at zero temperature, see e.g. \cite{Wald:1997qp}.

By now, several experimental groups include black holes into their
search for physics beyond the Standard Model. For detailed studies
of the experimental signatures, {\sc PYTHIA} 6.2 \cite{PYTHIA} has
been coupled to {\sc CHARYBDIS} \cite{CHARYBDIS} creating an event
generator allowing for the simulation of black hole events and data
reconstruction from the decay products. Previous analysis within
this framework are summarized in Refs. \cite{Atlas,BHex}. Ideally,
the energy distribution of the decay products allows a determination
of the temperature (by fitting the energy spectrum to the predicted
shape) as well as of the total mass of the object (by summing up all
energies). This then allows to reconstruct the scale $M_{\rm f}$ and
the number of extra dimensions.

These analysis however, have so far omitted the possibility of a
stable black hole remnant but assume instead a final decay into some
few particles, whose number is treated as a free parameter ranging
from $2-5$.

In the following we will examine the possibility that a stable black
hole remnant of about Planck mass is left with use of the {\sc
PYTHIA} 6.2/{\sc CHARYBDIS} event generator package.

\section{Black Hole Remnants}
\label{BHr}

The final fate of black holes is an unresolved subject of ongoing
research. The last stages of the evaporation process are closely
connected to the information loss puzzle. The black hole emits
thermal radiation, whose sole property is the temperature,
regardless of the initial state of the collapsing matter. So, if a
black hole completely decays into statistically distributed
particles, unitarity can be violated. This happens when the initial
state is a pure quantum state and then evolves into a mixed state
\cite{Novi,Hawk82,Preskill}.

When one tries to avoid the information loss problem two possibilities are left.
The information
is regained by some unknown mechanism or a stable black hole remnant is formed
which keeps the
information. Besides the fact that it is unclear in which way the information should escape the horizon
\cite{escape} there are several other arguments for
black hole remnants \cite{relics}:
\begin{itemize}
\item
The uncertainty relation: The Schwarzschild radius of a black hole with Planck mass
is of the order  of the Planck length. Since the Planck length is the wavelength corresponding to a particle of
Planck mass, a problem arises when the mass of the black hole drops below Planck mass.
Then one has trapped a mass inside a volume which is smaller than allowed by the uncertainty
principle \cite{39}. To avoid this problem, Zel'dovich has proposed that black holes with masses below
Planck mass should be associated with stable elementary particles \cite{40}.
Also, the occurrence of black hole remnants within the framework of a generalized uncertainty principle
has been investigated in \cite{Adler:2001vs,BHwithGUP}.

\item
Corrections to the Lagrangian: The introduction of additional terms,
which are quadratic in the curvature, yields a decrease of the
evaporation temperature towards zero \cite{Barrow,Whitt}. This holds
also for extra dimensional scenarios \cite{my2} and is supported  by
calculations in the low energy limit of string theory
\cite{Callan,Stringy}. The production of TeV-scale black holes in
the presence of Lovelock higher-curvature terms has been examined in
\cite{Rizzo:2005fz} and it was found that these black holes can
become thermodynamically stable since their evaporation takes an
infinite amount of time.

\item
Further reasons for the existence of remnants have been suggested to be black holes with axionic charge
\cite{axionic}, the modification of the Hawking temperature due to quantum hair \cite{hair} or magnetic
monopoles \cite{magn}. Coupling of a dilaton field to gravity also yields remnants, with detailed
features depending on the dimension of space-time \cite{dilaton1,dilaton2}.

\item One might also see the
arising necessity for remnant formation by applying the geometrical analogy to the black hole
and quantize the radiation into wavelengths that fit on the surface, i.e. the horizon \cite{Hossenfelder:2003dy}.
The smaller
the size of the black hole, the smaller the largest possible wavelength and the larger the
smallest possible energy quantum that can be emitted. Should the energy of the lowest energy
level already exceed the total mass of the black hole, then no further emission is possible.
Not surprisingly, this equality happens close to the Planck scale and results in the
formation of a stable remnant.
\end{itemize}

Of course these remnants, which in various context have also been named
Maximons, Friedmons, Cornucopions, Planckons or
Informons, are not a miraculous remedy but bring some new problems along. Such is e.g.
the necessity for an infinite number of states which allows the unbounded information
content inherited from the initial state.

\section{Signatures of Black Hole Remnants}
\label{signatures}

We now attempt to construct a numerically applicable model for
modifications of the black hole's temperature in order to simulate
the formation of a black hole remnant. Though the proposals of
remnant formation in the literature are build on various different
theoretical approaches, they have in common that the temperature of
the black hole drops to zero already at a finite black hole mass. We
will denote the mass associated with this finite remnant size with
$M_{\rm R}$ and make the reasonable identification $M_{\rm R} =
M_{\rm min}$. Instead of deriving such a minimal mass within the
frame of a specific model, we aim in this work to parametrize its
consequences for high energy collisions.

For our purposes, we will assume that we are dealing with a theory of modified gravity which
results in a remnant mass and parametrize the deviations of the entropy $S(M)$. This entropy
now might differ from the Hawking-entropy by correction terms in $M_{\rm R}/M$. For black
hole masses  $M$ much larger than $M_{\rm R}$  we require to reproduce the standard result. The
expansion then reads
\begin{equation} \label{entropy}
S(M) = {\cal{A}}_{(d+3)} M_f^{d+2} \left[ a_0 + a_1 \left(\frac{M_{\rm R}}{M}\right) + a_2
\left(\frac{M_{\rm R}}{M}\right)^2 + \dots\right]
\end{equation}
with dimensionless coefficients $a_i$ depending on the specific
model (see e.g. \cite{Adler:2001vs,Barrow,Stringy,dilaton2}). As
defined in Eq. (\ref{oberfl}), ${\cal{A}}$ is the surface of the
black hole and a function of $M$. For the standard scenario one has
\begin{equation}
 a_0 = \frac{d+1}{d+2} \frac{2 \pi}{\Omega_{(d+3)}}\quad, \quad  a_{i>1} = 0\quad. \label{coeffstandard}
\end{equation}
Note that in general
\begin{equation}
S_0= S(M=M_{\rm R})
\end{equation}
will differ from the unmodified black hole entropy since the Schwarzschild-radius can be modified.

It should be understood that an underlying theory of modified
gravity will allow to compute $M_{\rm R}=M_{\rm R}(a_i)$ explicitly
from the initially present parameters. This specific form of these
relations however, depends on the ansatz. We will instead treat
$M_{\rm R}$ as the most important input parameter. Though the
coefficients $a_i$ in principle modify the properties of the black
hole's evaporation, the dominating influence will come from the
existence of a remnant mass itself, making the $a_i$ hard to extract
from the observables.

To make this point clear, let us have a closer look at the
evaporation rate of the black hole by assuming a remnant mass. Note,
that the Hawking-evaporation law can not be applied towards masses
that are comparable to the energy of the black hole because the
emission of the particle will have a non-negligible back reaction.
In this case, the black hole can no longer be treated in the micro
canonical ensemble but instead, the emitted particles have to be
added to the system, allowing for a loss of energy into the
surrounding of the black hole.  Otherwise, an application of the
Hawking-evaporation down to small masses comparable to the
temperature of the black hole, would yield the unphysical result
that the evaporation rate diverges because one has neglected that
the emitted quanta lower the mass of the black hole.

This problem can be appropriately addressed by including the
back reaction of the emitted quanta as has been derived in
\cite{Page,backreaction,Parikh:1999mf}. It is found that in the
regime of interest here, when $M$ is of order $M_{\rm f}$, the
emission rate for a single particle micro state is modified and given
by the change of the black hole's entropy
\begin{equation} \label{nsingle}
n(\omega) =  \frac{\exp[S(M-\omega)]}{\exp[S(M)]}\quad.
\end{equation}
If the average energy of the emitted particles is much smaller than
$M$, as will be the case for $M \gg M_{\rm f}$, one can make the
approximation \beqn \label{approxsmallo} S(M) - S(M-\omega) \approx
\frac{\partial S}{\partial M} \omega =  \frac{\omega}{T} \eeqn
which, inserted in Eq.(\ref{nsingle}) reproduces the familiar
relation. The single particle distribution can be understood by
interpreting the occupation of states as arising from a tunnelling
probability \cite{Parikh:1999mf,Medved:2005yf}. From the single
particle number density (Eq. \ref{nsingle}) we obtain the average
particle density by counting the multi particle states according to
their statistics
\begin{equation} \label{nmulti}
n(\omega) = \left( \exp[S(M)-S(M-\omega)] + s \right)^{-1} \quad,
\end{equation}
where
\beqn
s &=& ~~~1 \quad \mbox{for Fermi-Dirac statistic} \nonumber \\
s &=& ~~~0 \quad \mbox{for Boltzmann statistic} \nonumber \\
s &=& -1 \quad \mbox{for Bose-Einstein statistic} \quad,
\eeqn
and $\omega \leq M-M_{\rm R}$, such that nothing can be
emitted that lowers the energy below the remnant mass. Note, that this number density will
assure that the remnant is formed even if the time variation of the black hole's temperature (or its
mass respectively) is not taken into account.

For the spectral energy density we then use this particle
spectrum and integrate over the momentum space. Since
we are concerned with particles of the Standard Model which are bound to the 3-brane, their
momentum space is the usual 3-dimensional one.
This yields
\begin{eqnarray}
\varepsilon = \frac{\Omega_{(3)}}{(2 \pi)^3}\zeta(4)
\int_0^{M-M_{\rm R}} \hspace*{-2mm}  \frac{\omega^3 ~{\mathrm d}\omega }{{\rm exp}[S(M)-S(M-\omega)] + s}  \quad.
\end{eqnarray}
From this, we obtain the evaporation rate with the Stefan-Boltzmann law to
\begin{eqnarray} \label{mdoteq}
\frac{{\mathrm d}M}{{\mathrm d}t} = \frac{\Omega_{(3)}^2}{(2\pi)^{3}} R_H^{2}\zeta(4)
\int_0^{M-M_{\rm R}} \hspace*{-2mm}
\frac{\omega^3 ~{\mathrm d}\omega }{{\rm exp}[S(M)-S(M-\omega)] + s}   \quad .
  \label{dmdtrel}
\end{eqnarray}
Since we are dealing with emitted particles bound to the brane, the surface through which the
flux disperses is the $2$-dimensional intersection of the black hole's horizon with the brane.

\begin{figure}
\vspace*{-1cm}
\hspace*{-1cm}
\epsfig{figure=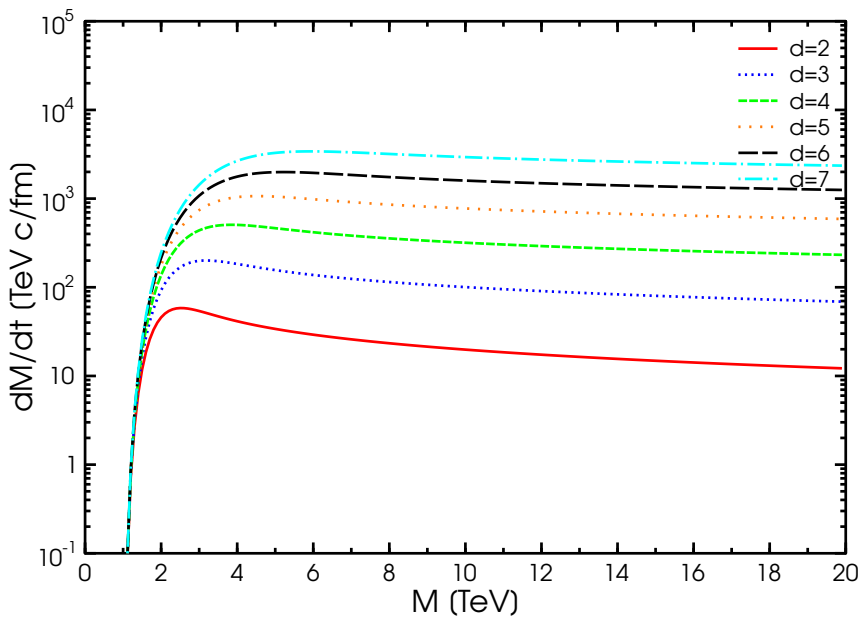,width=14.5cm}

\vspace*{-1cm} \caption{The evaporation rate Eq. (\protect\ref{mdoteq}) for
various $d$ for $M_{\rm R} = M_{\rm f} =1$~TeV and the standard
entropy, i.e. the parameter set Eq. (\protect\ref{coeffstandard}). Here,
Boltzmann- statistic was used. \label{fig1}}
\end{figure}

\begin{figure}
\vspace*{-1cm}
\hspace*{-1cm}
\epsfig{figure=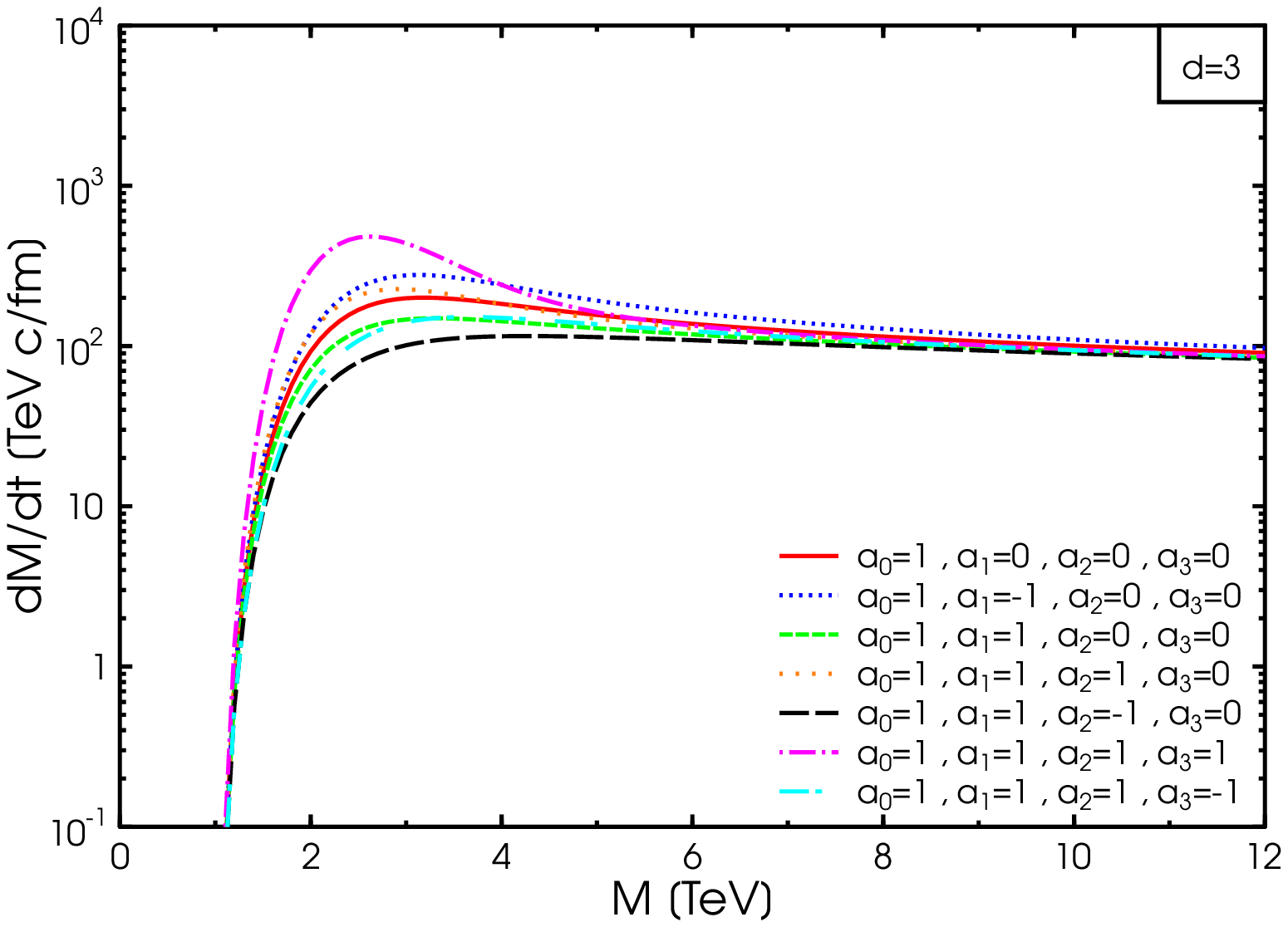,width=14.5cm}

\vspace*{-1cm} \caption{The evaporation rate for the black hole with
$M_{\rm R}=M_{\rm f}=1$~TeV and $d=3$ for various parameters $a_i$.
Here,  Boltzmann-statistic was used. \label{fig2}}
\end{figure}

\begin{figure}
\vspace*{-1cm}
\hspace*{-1cm}
\epsfig{figure=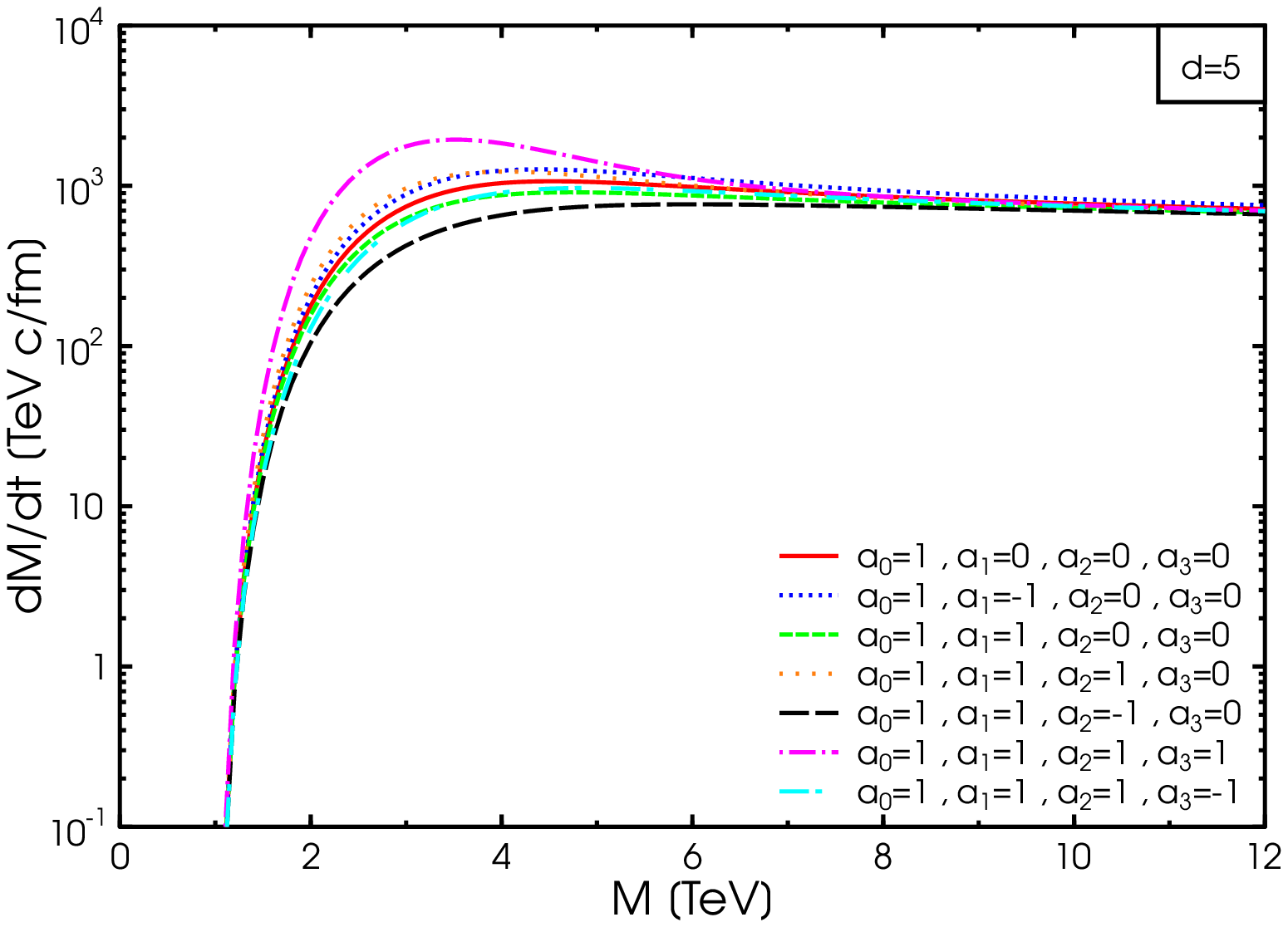,width=14.5cm}

\vspace*{-1cm} \caption{The evaporation rate for the black hole with
$M_{\rm R}=M_{\rm f}=1$~TeV and $d=5$ for various parameters $a_i$.
Here, Boltzmann-statistic was used. \label{fig3}}
\end{figure}

Inserting the modified entropy Eq. (\ref{entropy}) into the derived
expression Eq. (\ref{dmdtrel}), one sees that the evaporation rate
depends not only on $M_{\rm R}$ but in addition on the free
parameters $a_i$. However, for large $M$ the standard scenario is
reproduced and we can apply the canonical ensemble. E.g. for the
Fermi-Dirac statistic one obtains
\begin{eqnarray} \label{largem}
\frac{{\mathrm d}M}{{\mathrm d}t} = \frac{\Omega_{(3)}^2}{(2\pi)^{3}} R_H^{2} \zeta(4) \Gamma(4) T^4
\quad\mbox{for}\quad M \gg M_{\rm R} \quad.
\end{eqnarray}
Whereas for $M/M_{\rm R} \to 1$, the dominant contribution from the
integrand in Eq. (\ref{dmdtrel}) comes from the factor $\omega^{3}$
and the evaporation rate will increase with a power law. The slope
of this increase will depend on $S_0$. From this qualitative
analysis, we can already conclude that  the coefficients $a_i$ will
influence the black hole's evaporation  only in the intermediate
mass range noticeably. If we assume the coefficients to be in a
reasonable range -- i.e. each $a_i$  is of order $1$ or less and the
coefficient $a_{i+1}$ is smaller\footnote{From naturalness, one
would expect the coefficients to become smaller with increasing $i$
by at least one order of magnitude see e.g. \cite{Rizzo:2005fz}.}
than the coefficient $a_i$ and the series breaks off at a finite $i$
-- then the deviations from the standard evaporation are negligible
as is demonstrated in Figs. \ref{fig1}, \ref{fig2} and \ref{fig3}.

Figure \ref{fig1} shows the evaporation rate Eq. (\ref{dmdtrel}) for various $d$ with the standard
parameters (\ref{coeffstandard}). Figure \ref{fig2} and \ref{fig3} show various choices of parameters
for $d=3$  and $d=5$ as examples. Note, that setting $a_3$ to $1$ is already in a very extreme
range since a natural value was several orders of magnitude smaller: $a_3 \leq 10^{-3}$ (in this case the
deviations would not be visible in the plot).
For our further numerical treatment, we have included the possibility to vary the $a_i$ but one
might already at this point expect them not to have any influence on the characteristics of
the black hole's evaporation except for a slight change in the temperature-mass relation.

\begin{figure}
\vspace*{-1cm}
\hspace*{-1cm}
\epsfig{figure=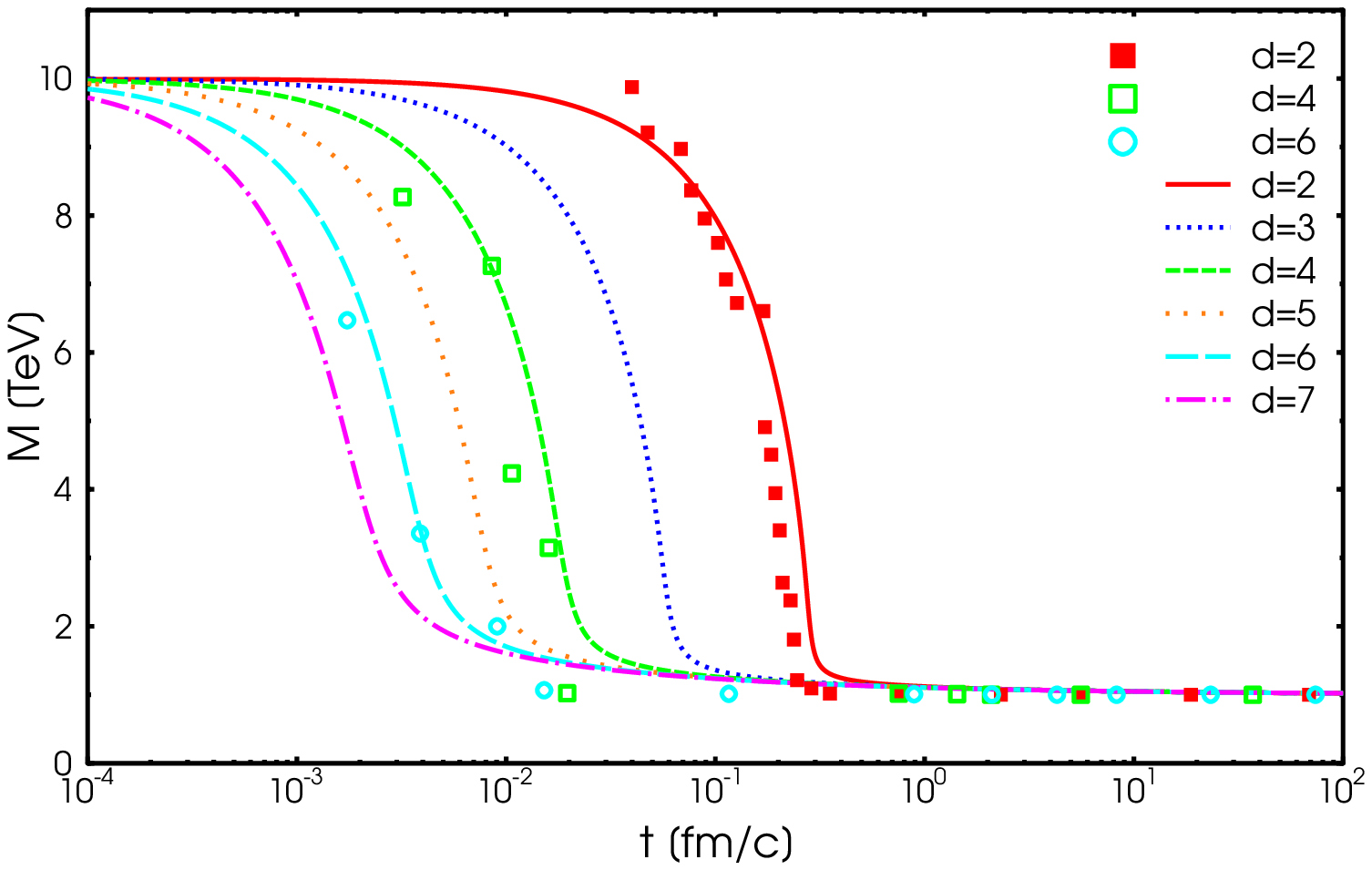,width=14.5cm}

\vspace*{-1cm}
\caption{The mass evolution for a black hole of initial mass $M=10$~TeV and
various $d$. Here, we set $M_{\rm R} = M_{\rm f} =1$~TeV. The full lines
show the analytical calculation. The numerical results are shown as symbols.
Note that each numerical example shows a single event only.
\label{fig4}}
\end{figure}

From the evaporation rate Eq.(\ref{dmdtrel}) one obtains by integration the mass evolution $M(t)$
of the black hole. This is shown for the continuous mass case in Figure \ref{fig4}. For a realistic
scenario one has to take into account that the mass loss will proceed by steps by radiation
into the various particles of the Standard Model.

\section*{Results}

We have included the evaporation rate, parametrized according to the
previous section, into the black hole event generator {\sc CHARYBDIS}
and examined the occurring observables within the {\sc PYTHIA}
environment. Since these black hole remnants are stable, they are of
special interest as they are available for close investigations.
Especially those remnants carrying an electric charge offer exciting
possibilities as  investigated in \cite{second}.

It has also been shown in \cite{second} that no naked singularities
have to be expected for reasonably charged black holes and that the
modification of the Hawking radiation due to the electric charge can
be neglected for the parameter ranges one expects at the {\sc LHC}.
This means in particular that the interaction of emitted charged
particles with the black hole does not noticeably modify the emission
probability. Although there might be uncertainties in the low energy
limit where QED or QCD interactions might have unknown consequences
for the processes at the horizon.

The formation of a remnant indeed solves a (technical) problem
occuring within the treatment of a final decay: it might in
principle have happened that during its evaporation process, the
black hole has emitted mostly electrically charged particles and
ended up with an electric charge of order ten. In such a state, it
would then be impossible for the black hole to decay into less than
ten particles of the SM, whereas the standard implementation allows
only a decay into a maximum of 5 particles.

Therefore, in the original numerical treatment, the process of
Hawking radiation has before been assumed to minimize the charge of
the evaporating hole in each emission step. In such a way, it was
assured that the object always had a small enough charge to enable
the final decay in $\leq 5 $ particles without any violation of
conservation laws. This situation changes if the remnant is allowed
to keep the electric charge. In the here presented analysis, the
assumption of charge minimization has therefore been dropped as it
is no longer necessary. However, we want to stress, that the in- or
exclusion of charge minimization does not modify the observables
investigated\footnote{The differences in the finally observable charged
particle distributions from the black hole decay are changed by
less than 5\% compared to the charge minimization setting.}.

When attempting to investigate slowly decaying objects, one might be
concerned whether these decay in the collision region or might be
able to leave the detector, thereby still emitting radiation. As
shown for the continuous case in Fig. \ref{fig4}, the average energy
of the emitted particles drops below an observable range within a
$10$~fm radius. Even if one takes into account the large
$\gamma$-factor, the black hole will have shrunken to remnant-mass
safely in the detector region. This is shown for a sample of
simulated events in Fig. \ref{fig4} (symbols) which displays the mass
evolution of these collider produced black holes. Here, we estimated
the time, $t$, for the stochastic emission  of a quanta of energy
$E$ to be $1/E$. This numerical result agrees very well with the
expectations from the continuous case.
\begin{figure}[t!]

\vspace*{-1.0cm} \hspace*{-1cm} \epsfig{figure=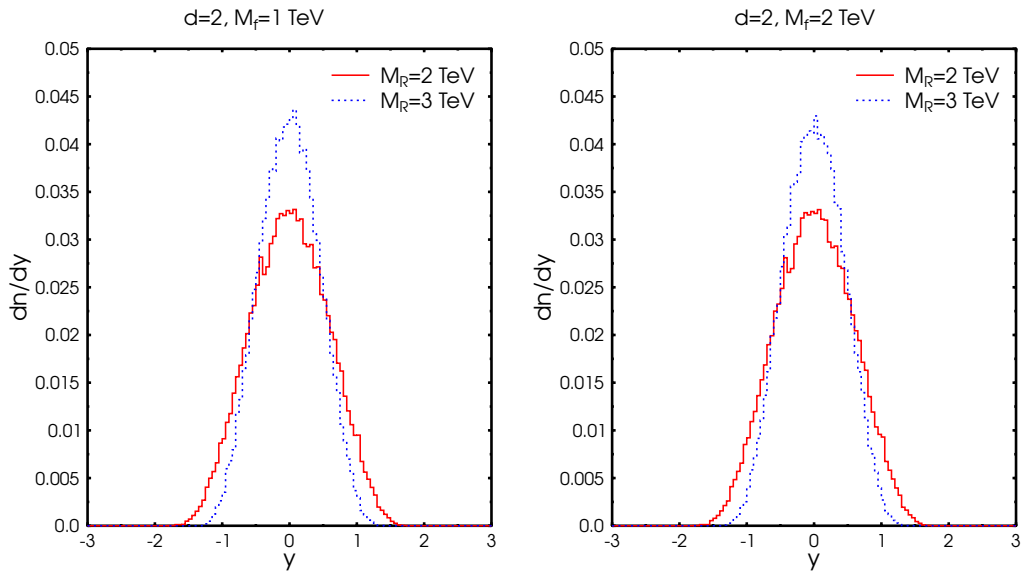,width=14cm}

\caption{Rapidity distribution of the black hole remnants in pp
interactions at $\sqrt s=14$~TeV for $d=2$. The curves for different
number of extra dimensions $d$ differ from the depicted ones by less
than 5\% and are not shown. \label{fig6}}
\end{figure}
\begin{figure}[h!]

\vspace*{0.5cm} \hspace*{-1cm}
\epsfig{figure=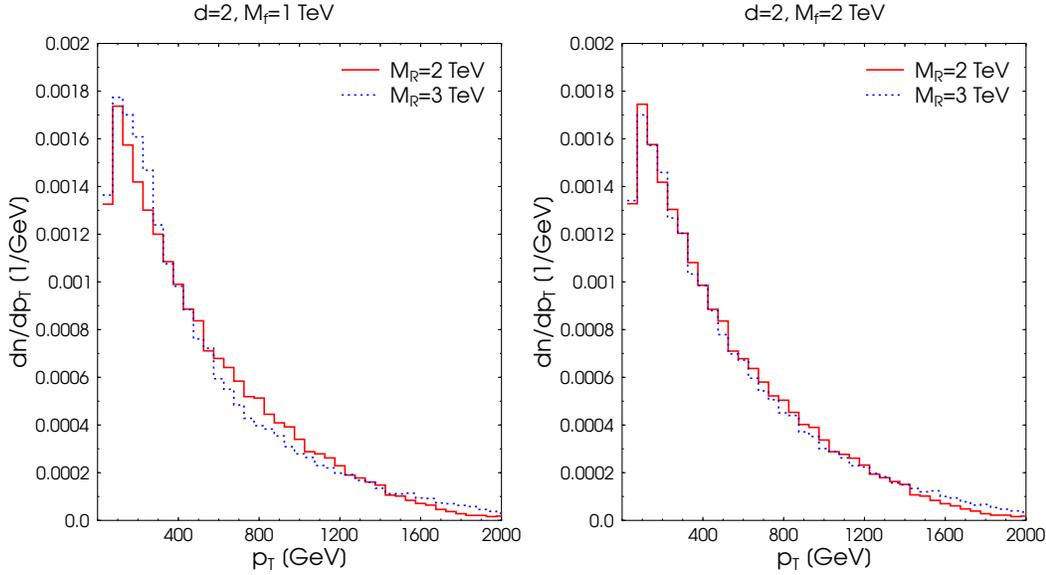,width=14cm}

\caption{Transverse momentum distribution of the black hole remnants
in pp interactions at $\sqrt s=14$~TeV. \label{fig7}}
\end{figure}

\begin{figure}[t!]

\vspace*{-1.2cm} \hspace*{-1cm} \epsfig{figure=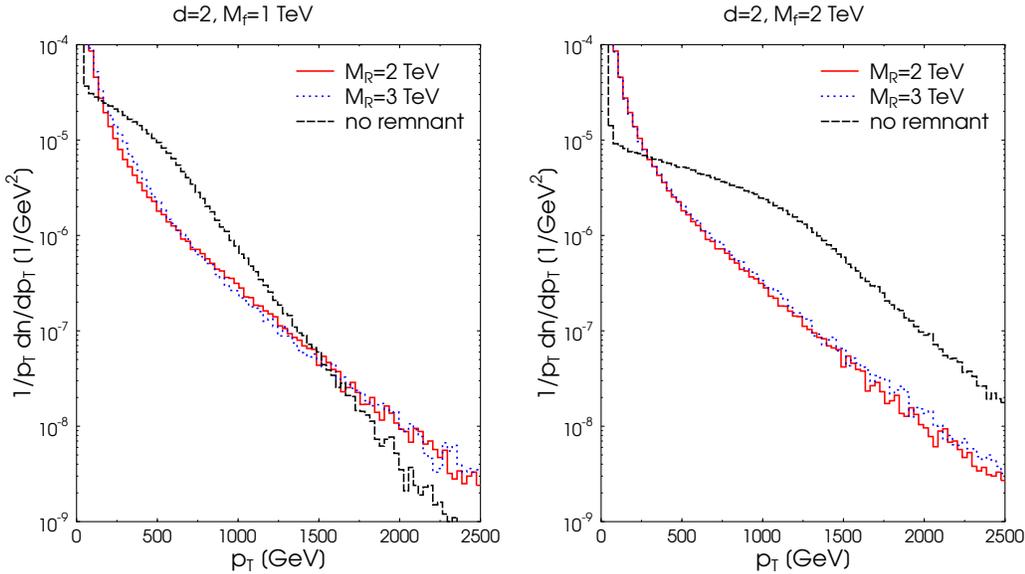,width=14cm}

\caption{Transverse momentum distribution of  initially emitted
particles (i.e. before the fragmentation of the emitted partons)
with final (two-body) decay in contrast to the formation of a black
hole remnant. \label{fig8}}
\end{figure}
\begin{figure}[h!]

\vspace*{1.0cm} \hspace*{-1cm} \epsfig{figure=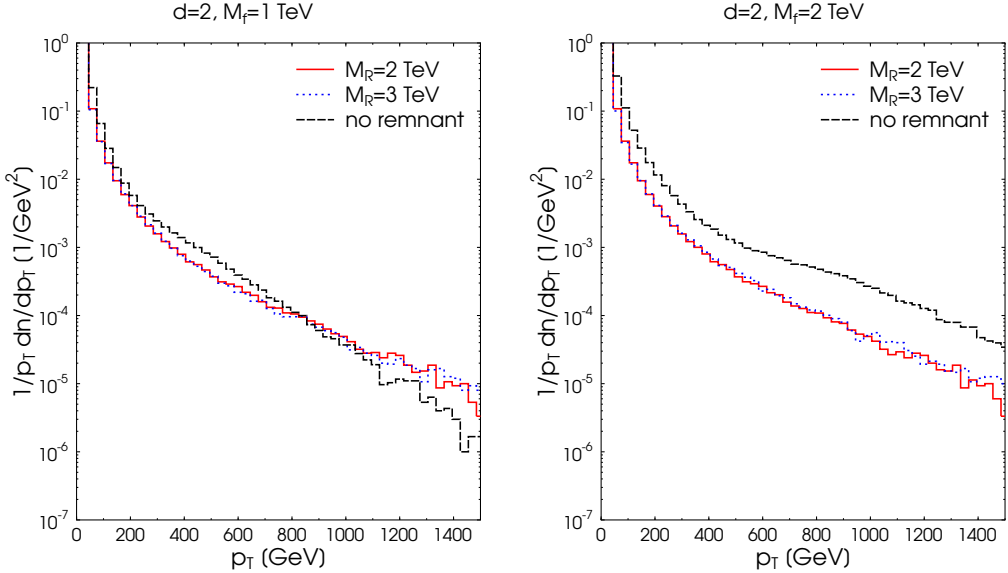,width=14cm}

\caption{Transverse momentum distribution after fragmentation with
final (two-body) decay in contrast to the formation of a black hole
remnant. \label{fig9}}
\end{figure}

\begin{figure}

\vspace*{-1.0cm} \hspace*{-1cm}
\epsfig{figure=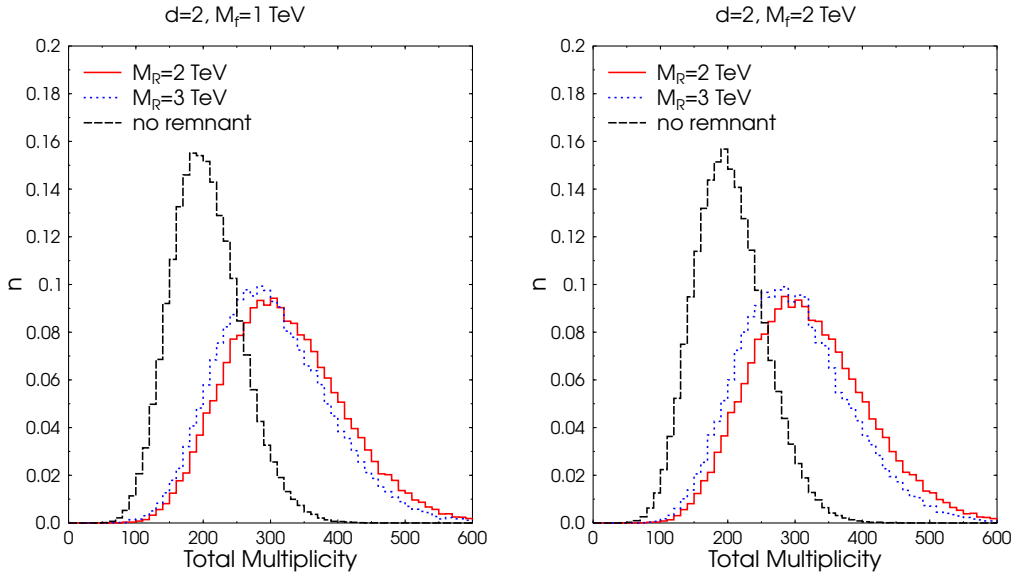,width=14cm}

\caption{Total  multiplicity with final (two-body) decay in contrast
to the formation of a black hole remnant for $d=2$. \label{fig10}}
\end{figure}

\begin{figure}

\vspace*{-0.0cm} \hspace*{-1cm}
\epsfig{figure=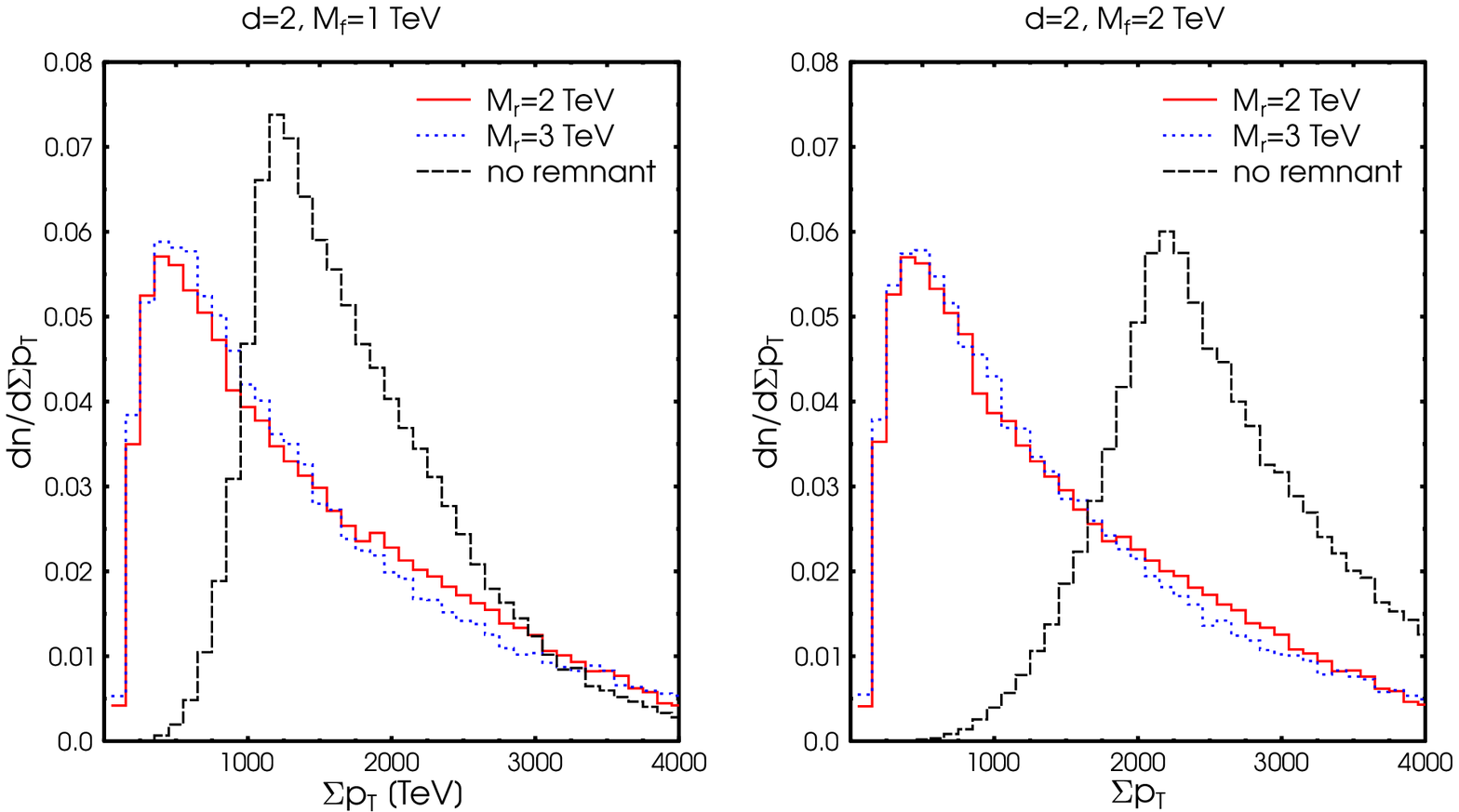,width=14cm}

\caption{The total sum of the transverse momenta of the decay
products. \label{fig11}}
\end{figure}

To understand the fast convergence of the black hole mass, recall
the spectral energy density which enters in Eq.(\ref{dmdtrel}) and
which dictates the distribution of the emitted particles. Even
though the spectrum is no longer an exactly Planckian, it still
retains a maximum at energies $\sim 1/T$. If the black hole's mass
decreases, the emission of the high energetic end of the spectrum is
no longer possible. For masses close to the Planck scale, the
spectrum has a maximum at the largest possible energies that can be
emitted. Thus, the black hole has a high probability to emit its
remaining energy in the next emission process. However,
theoretically, the equilibrium time goes to infinity (because the
evaporation rate falls to zero, see Fig. \ref{fig1}) and the black
hole will emit an arbitrary amount of very soft photons. For practical
purposes, we cut off\footnote{The emission of objects carrying color
charge is disabled after the maximally possible energy drops below
the mass of the lightest meson, i.e. the pion.} the evaporation when
the black hole reached the mass $M_{\rm R} + 0.1$~GeV.

Figure \ref{fig6} shows the rapidity of the produced black hole
remnants in a proton-proton collision at $\sqrt{s}=14$~TeV.
 All plots are for $d=2$ since a higher number of
extra dimensions leads to variations of less than 5\%.
The reader should be aware that the present numerical studies assume the production
of one black hole in every event. To obtain the absolute cross sections
the calculated yields have to be multiplied by the black hole production cross section
$\sigma({\rm pp}\rightarrow BH)$. Due to the uncertainties in the absolute production
cross section of black holes we have taken this factor explicitely out.
For the
present examination we have initialized a sample of 50,000 events.
The black hole remnants are strongly peaked around central rapidities,
making them potentially accessible to the CMS and ATLAS experiments.
In Figure \ref{fig7} we show the distribution of the produced black
hole remnants as a function of the transverse momentum.

Figure \ref{fig8} shows the transverse momentum, $p_T$, of the decay
products as it results from the modified multi particle number
density Eq. (\ref{nmulti}) before fragmentation. Figure \ref{fig9}
shows the $p_T$-spectrum after fragmentation. In both cases, one
clearly sees the additional contribution from the final decay which
causes a bump in the spectrum which is absent in the case of a
remnant formation. After fragmentation, this bump is slightly washed
out but still present. However, from the rapidity distribution and
the fact that the black hole event is spherical, a part of the high
$p_T$-particles will be at large $y$ and thus be not available in
the detector. We therefore want to mention that one has to include
the experimental acceptance in detail if one wants to compare to
experimental observables.

 Figure \ref{fig10} shows the total
multiplicities of the event. When a black hole remnant is formed,
the multiplicity is increased due to the additional low energetic
particles that are emitted in the late stages instead of a final
decay with $2-5$ particles. Note that this multiplicity increase is
not an effect of the remnant formation itself, but stems from the treatment
of the decay in the micro-canonical ensemble used in the present
calculation. I.e. the black hole evaporates a larger amount of
particles with lower average energy.

Figure \ref{fig11}  shows the sum over the transverse momenta of the
black holes' decay products. To interpret this observable one might
think of the black hole event as a multi-jet with total $\Sigma p_T$.
As is evident, the formation of a remnant lowers the total $\Sigma p_T$ by
about $M_{\rm R}$. This also means, that the signatures of the black
hole as previously analyzed are dominated by the daubtful final decay and
not by the Hawking phase.
It is interesting to note that the dependence on $M_{\rm f}$ is
dominated by the dependence on $M_{\rm R}$, making the remnant mass the
primary observable, leading to an increase in the missing energy.

\section{Conclusion}
\label{Concl}

We have parametrized the modifications to the black hole evaporation
arising from the presence of a remnant mass. The modified spectral
density is included in the numerical simulation for black hole
events. To give a specific example, we have examined the formation
of black hole remnants in proton proton collision at $\sqrt
s=14$~TeV and set it in contrast to a final decay of the black hole.

We predict a significant decrease of the total transverse momentum
of the black hole remnant events due to the absence of the final decay
particles. Even more, the multiplicity of the event is increased by
a factor $\sim 3/2$ arising from the micro-canonical treatment of
the evaporation process.

The formation of the black hole remnant results in a strong
modification of most predicted black hole signatures. However, remnant
formation itself leads to prominent experimental signatures (see
e.g. $\Sigma p_T$). This makes the search for black hole remnants
promising and experimentally accessible for the CMS and ATLAS
experiments.

\section*{Acknowledgements}

We thank Horst St\"ocker for helpful discussions. This work was
supported by NSF PHY/0301998 and {\sc DFG}. SH thanks the FIAS for
kind hospitality.

\end{document}